\font\bbf=cmbx12

\font\srm=cmr9
\font\sit=cmti9
\font\sbf=cmbx9

\
{\srm \hfill $\copyright$ 2000 The American Institute of Physics}
\bigskip
\centerline{\srm to be published in AIP Conference Proceedings of the}
\centerline{\srm Space Technology and Applications International Forum
(STAIF-2000)}\par
\centerline{\srm Conference on Enabling Technology and Required
Scientific Developments for Interstellar Missions}\par
\centerline{\srm January 30---February 3, 2000, Albuquerque, NM}\par
\vskip 0.2truein
\centerline{\bbf TOWARD AN INTERSTELLAR MISSION:}
\centerline{\bbf ZEROING IN ON THE ZERO-POINT-FIELD
INERTIA RESONANCE}
\bigskip
\centerline{\bbf Bernhard Haisch$^1$ and Alfonso Rueda$^2$}
\bigskip
\centerline{\sit $^1$Solar \& Astrophysics Laboratory, Lockheed Martin, L9-41, B252, 3251
Hanover St., Palo Alto, CA 94304}
\centerline{\sit $^2$Dept. of  Electrical Eng. and Dept. of Physics \& Astronomy,
California State Univ., Long
Beach, CA 90840}
\bigskip
\centerline{\sit $^1$haisch@starspot.com \ \ \ $^2$arueda@csulb.edu}
\bigskip
\bigskip

{\parindent 0.2truein \narrower \noindent
{\sbf Abstract.}
While still an admittedly remote possibility, the concept of an interstellar mission has
become a legitimate topic for scientific discussion as evidenced by several recent NASA
activities and programs. One approach is to extrapolate present-day technologies by orders
of magnitude; the other is to find new regimes in physics and to search for possible new
laws of physics. Recent work on the zero-point field (ZPF), or electromagnetic quantum
vacuum, is promising in regard to the latter, especially concerning the possibility that
the inertia of matter may, at least in part, be attributed to interaction between the
quarks and electrons in matter and the ZPF. A NASA-funded study (independent of the BPP
program) of this concept has been underway since 1996 at the Lockheed Martin Advanced
Technology Center in Palo Alto and the California State University at Long Beach. We
report on a new development resulting from this effort: that for the specific case of the
electron, a resonance for the inertia-generating process at the Compton frequency would
simultaneously explain both the inertial mass of the electron and the de Broglie
wavelength of a moving electron as first measured by Davisson and Germer in 1927. This
line of investigation is leading to very suggestive connections between electrodynamics,
inertia, gravitation and the wave nature of matter.

}

\bigskip\centerline{\bbf BACKGROUND}
\medskip\noindent
Although at this time there are no known or plausible technologies which would make
interstellar travel possible, the concept of an interstellar mission has recently started
to become a legitimate topic for scientific discussion. In July 1996 NASA formally
established a  Breakthrough Propulsion Physics (BPP) Program. The initiation of a program 
operating under the auspices of NASA provides a forum for ideas which are required to be
both {\it visionary} and {\it credible}. This approach was designed by program manager Marc
Millis to provide a valuable and necessary filter for ideas and their proponents. The
first BPP activity was a workshop held in August 1997 to carry out an initial survey of
concepts; this is now available as a NASA conference proceedings (NASA/CP---1999-208694).
Following additional funding, the BPP program issued a NASA Research Announcement (NRA) in
November 1998 soliciting proposals. Out of 60 submissions, six studies were selected for
funding.

\medskip\noindent
Three other events relevant to interstellar exploration but not directly connected with the
BPP program have also taken place. In February 1998, the Marshall Space Flight Center
(MSFC) hosted a four-day workshop on ``Physics for the Third Millenium'' directed by Ron
Koczor of the MSFC Space Science Laboratory. In July 1998 a four-day workshop on ``Robotic
Interstellar Exploration in the Next Century'' sponsored by the Advanced Concepts Office
at JPL and the NASA Office of Space Science was held at Caltech. In September 1998 an
Independent Review Panel was convened to assess the NASA Space Transportation Research
Program (STRP). The STRP supports such areas of investigation as fission- and fusion-based
advanced propulsion and even a serious attempt to replicate a claimed ``Delta-g Gravity
Modification Experiment.'' A report was issued in January 1999 entitled: ``Ad Astra per
Aspera: Reaching for the Stars.''

\medskip\noindent
The report had the following to say about the Delta-g Gravity
Modification Experiment:

\medskip\noindent

{\baselineskip=12pt \narrower

An experimental demonstration of gravity modification,
regardless of how minute, would be of extraordinary significance. The Delta-g experiments
now being carried out at MSFC are an attempt to duplicate a claimed anomalous weight loss
of up to two percent in objects of various compositions suspended above a rotating 12-inch
diameter Type II ceramic superconductor. This apparent phenomenon was discovered by
accident during superconducting experiments by a Russian scientist, Dr. Eugene Podkletnov,
then working at Tampere University in Finland. Initial results of the MSFC replication
have been published in Physica C, 281, 260, 1997. As of November 1998 the group, led by
David Noever and Ron Koczor, had made a 12-inch YBCO disk that survived pressing and heat
treating in one piece. This is now being characterized and cut up to do mechanical
testing. The next step is to make a 2-layer disk of the sort used by Podkletnov.\par The
review committee was impressed with the high quality of the researchers and the careful
and methodical approach. We believe that this research is a prime candidate for continued
STR support and urge that funding is adequate to permit a definitive replication to be
carried to completion.''

}

\medskip\noindent
In all of these activities (in which the first author participated) it became clear
that there are two approaches to conceptualizing an interstellar mission. The first is to
extrapolate certain relevant present-day technologies by orders of magnitude and then see
what possibilities might emerge. One example --- discussed at the Caltech workshop ---
would be a craft propelled by a laser-pushed lightsail\dots but this would require a 1000
km-diameter sail and a 10 km-diameter lens having an open-loop pointing capability of
$10^{-5}$ arcsec (given a feedback time of years): quite a formidable challenge!
\medskip\noindent
Another ``known technology pushed to the limit'' example is based on production and storage
of huge amounts of anti-matter. This makes a good example of overwhelming technical
difficulties in pushing present-day technology.
Based on energy arguments alone (i.e. ignoring the issue of specific impulse) one can
achieve a speed of 0.1$c$ by annihilating 0.5 percent of the mass of a starship; this is
simply calculated by equating the final kinetic energy of the starship, $m_sv^2/2$, where
$v=0.1c$, to the rest energy of the propellant, $m_pc^2$, a good approximation in this
regime. It would take an equal amount of energy to stop, and similarly for the return to
Earth. Under perfectly ideal conditions then, one percent of the mass of the starship in
antimatter and one percent in ordinary matter would, in principle, suffice as propellant.
One hundred percent efficiency for conversion of rest mass energy into kinetic energy is
out of the question. Let us take an optimistic 10 percent. This means that for a starship
as modest as the space shuttle in size --- about 100 tons, hardly adequate for a
century-long out and back mission to Alpha Centauri --- one would require 10 tons of
antimatter.

\medskip\noindent
The manufacture of antiprotons is extremely inefficient. 
Techniques for creating antiprotons at CERN require approximately two and one-half
million protons each accelerated to an energy of 26 GeV to create a single
antiproton. This amounts to an energy efficiency of $3 \times 10^{-8}$. This is further
reduced by a factor of 20 or so for the efficiency of the proton accelerator, leaving a
net efficiency of perhaps $1.5 \times 10^{-9}$, i.e. about one part in a billion! At a
cost of 5 cents per kilowatt-hour of electricity the cost of 10 tons of antiprotons
would be $1.4 \times 10^{21}$ dollars. A good way to imagine this is to say that
it represents the total current U. S. federal budget (app. \$1.2 trillion per year) spent
every year for the next 1.2 billion years (cf. M. R. LaPointe, NASA SBIR Phase I Final
Report  for contract NAS8-98109). Other technological extrapolations for propulsion
discussed in the various workshops and reviews suffer from similar tremendous
order-of-magnitude problems. Building a starship based on extrapolation of known
technology might be likened to insisting on building some kind of sailing ship capable of
crossing the Atlantic in 6 hours, when one really has to discover flight to do that.

\medskip\noindent
The second approach is to try to find new regimes in physics or perhaps even new laws of
physics. When Alcubierre (1994) published his article, ``The Warp Drive: Hyper-fast Travel
within General Relativity,'' this aroused considerable enthusiasm since it demonstrated
that, in principle, general relativity allowed for local metric modifications
resulting in expansion of space faster than the speed-of-light. Indeed, as is well known in
cosmology, there is no speed limit to the expansion of space itself; in conventional
inflationary big bang theory there must be regions of the Universe beyond our event horizon
whose Hubble speed (relative to us) is greater than $c$ and this in no way conflicts with
special relativity. Relativity merely forbids motion {\it through} space faster than $c$.
Alcubierre demonstrated that the mathematics of general relativity allowed for the
creation of what might be termed a ``bubble'' of ordinary flat (Minkowski) space --- in
which a starship could be situated --- that could surf, so to speak, at arbitrarily large
speeds on a spacetime distortion, a faster than light stretching of spacetime; this would
indeed be a warp drive. The Alcubierre bubble was soon burst, though, by Pfenning and Ford
(1997) who showed that it would require more energy than that of the entire Universe to
create the extremely warped space necessary. However recently Van den Broeck (1999) has
shown that this energy requirement can be reduced by 32 orders of magnitude via a slight
change in the Alcubierre geometry. While this still leaves us a long way from a feasible
interstellar technology, warping space or creating wormholes are physics possibilities
meriting further theoretical exploration.

\medskip\noindent
Another regime of ``new physics'' is in actuality almost a century old. The concept of an
electromagnetic zero-point field was developed by, among others, Planck, Einstein and
Nernst. If an energetic sea of electromagnetic fluctuations comprising the
electromagnetic quantum vacuum fills the Universe --- for reasons discussed below --- then
this suggests the possibility of generating propulsive forces or extracting energy
anywhere in space. Even more intriguing possibilities are opened by the linked proposed
concepts that gravitation and inertia may originate in the zero-point field. If both
gravitation and inertia are manifestations of the vacuum medium and in particular of its
electromagnetic component, the ZPF, they can be treated by the techniques of
electrodynamics, and perhaps they can be manipulated. The concept of gravity manipulation
has been a staple of science fiction, but in fact inertia manipulation would be even more
far reaching. As exciting as it would be to reduce the (gravitational) weight of a launch
vehicle to zero, this would merely set if free from the gravitational potential well of
the Earth and of the Sun. The problem of adding kinetic energy to reach high interstellar
velocities would remain\dots unless one can modify inertia. Modification of inertia would
(a) reduce energy requirements to attain a given velocity and (b) possibly allow greatly
enhanced accelerations. The latter would open many possibilities since it would be far more
efficient to have a perhaps enormously large acceleration device that never has to leave
the ground and needs to act over only a short distance to rapidly impart a huge impulse,
slingshotting a starship on its way. (We assume that the inertia of everything inside
the starship would be modified as well. For the time being we overlook the problem of
deceleration at the end of the journey, it being prudent to tackle only one apparent
impossibility at a time.)

\medskip\noindent
The concept of inertia modification may forever remain a {\it practical impossibility}.
However at the moment it has become a legitimate {\it theoretical possibility}. In the
following sections we summarize a recently developed theoretical connection between the ZPF
and inertia and report on the discovery that a specific resonance frequency is likely to
be involved. It is shown that such a resonance would
simultaneously offer an explanation for both the inertia of a particle and the
de Broglie wavelength of that particle in motion as first measured for electrons by
Davisson and Germer (1927).

\bigskip\centerline{\bbf THE ELECTROMAGNETIC ZERO-POINT FIELD}
\medskip\noindent
The necessary existence of an electromagnetic zero-point field can be shown from
consideration of  elementary quantum mechanics. The Heisenberg uncertainty relation tells
us that a harmonic oscillator must have a minimum energy of $\hbar \omega/2$, where
$\hbar$ is the Planck constant divided by $2\pi$ and $\omega$ is the oscillation frequency
in units of radians per second. (Expressed in cycles per second, Hz, this
minimum energy is
$h\nu/2$.) This is the zero-point energy of a harmonic oscillator, the derivation of which
is a standard example in many introductory quantum textbooks.

\medskip\noindent
The electromagnetic
field is subject to a similar quantization: this is done by ``the association of a
quantum-mechanical harmonic oscillator with each mode of the radiation field'' (cf.
Loudon 1983). The same 
$h \nu/2$ zero-point energy is found in each mode of the field, where a mode of the field
can be thought of as a plane wave specified by its frequency ($\nu$), directional
propagation vector ($\hat{\bf k}$), and one of two polarization states ($\sigma$).
Summing up over all 
plane waves one arrives at the zero-point energy density for the
electromagnetic field,
$$\rho_{ZP}(\nu) = \int_0^{\nu_c} {4 \pi h \nu^3 \over c^3} d\nu \ , \eqno(1)
$$
\medskip\noindent
where $\nu_c$ is a presumed high-frequency cutoff, often taken to be the Planck frequency,
$\nu_P = (c^5/G\hbar)^{1/2}=1.9 \times 10^{43}$ Hz. (See the appendix for a brief
discussion of the Planck frequency). With this assumed cutoff, the energy density becomes
the same (within a factor of
$2\pi^2$) as the maximum energy density that spacetime can sustain: with $\nu_c=\nu_P$,
the ZPF energy density is
$\rho_{ZP}=2 \pi^2 c^7/G^2
\hbar$.
This is on the order of $10^{116}$ ergs cm$^{-3}$ s$^{-1}$.
The term ``ZPE'' is often used to refer to this electromagnetic energy of the
zero-point fluctuations of the quantum vacuum. Note that the strong and weak interactions
also have associated zero-point energies. These should also contribute to inertia. Their
exact contributions remain to be determined since we have yet to
consider these in the present context. For now we restrict ourselves to the
electromagnetic contribution.

\medskip\noindent
Can one take seriously the concept that the entire Universe is filled with a background
sea of electromagnetic zero-point energy that is nearly 100 orders of magnitude beyond the
energy equivalent of ordinary matter density? The concept is inherently no more
unreasonable in modern physics that that of the vast Dirac sea
of negative energy anti-particles. Moreover the derivation of the zero-point energy from
the Heisenberg uncertainty relation and the counting of modes is so elementary that it
becomes convoluted to try to simply argue away the ZPF. The objection that most immediately
arises is a cosmological one: that the enormous energy density of the ZPF should, by
general relativity, curve the entire Universe into a ball orders of magnitude smaller than
the nucleus of an atom. Our contention is that the ZPF plays a key role in giving rise to
the inertia of matter. If that proves to be the case, the principle of equivalence will
require that the ZPF be involved in giving rise to gravitation. This at least puts the
spacetime-curvature objection in abeyance: in a self-consistent ZPF-based theory of inertia
and gravitation one can no longer naively attribute a spacetime curving property to the
energy density of the ZPF itself (Sakharov, 1968; Misner, Thorne and Wheeler, 1973;
Puthoff, 1989; Haisch and Rueda, 1997; Puthoff, 1999).

\medskip\noindent
One might try taking the position that the zero-point energy must be
merely a mathematical artifact of theory. It is sometimes argued, for
example, that the zero-point energy is merely equivalent to an 
arbitrary additive potential energy constant. Indeed, the potential
energy at the surface of the earth can take on any arbitrary value,
but the falling of an object clearly demonstrates the reality of a
potential energy field, the gradient of which is equal to a force.
No one would argue that there is no such thing as potential energy
simply because it has no well-defined absolute value. Similarly,
gradients of the zero-point energy manifest as measurable Casimir forces,
which indicates the reality of this sea of energy as well. 
Unlike the potential energy, however, the zero-point energy is
not a floating value with no intrinsically defined reference level.
On the contrary, the summation of modes tells us precisely how much
energy each mode must contribute to this field, and that energy density
must be present unless something else in nature conspires to cancel it.

\medskip\noindent
Another argument for the physical reality of zero-point fluctuations 
emerges from experiments in cavity quantum electrodynamics
involving suppression of spontaneous emission. As Haroche and Raimond (1993)
explain:

{\baselineskip=12pt \narrower

These experiments indicate a counterintuitive phenomenon that might be
called ``no-photon interference." In short, the cavity prevents an atom
from emitting a photon because that photon would have interfered 
destructively with itself had it ever existed. But this begs a 
philosophical question: How can the photon ``know," even before being
emitted, whether the cavity is the right or wrong size?

}

\noindent The answer is that spontaneous emission can be interpreted as 
stimulated emission by the ZPF, and that, as in the Casimir force
experiments, ZPF modes can be suppressed, resulting in no vacuum-%
stimulated emission, and hence no ``spontaneous" emission (McCrea, 1986).

\medskip\noindent
The Casimir force attributable to the ZPF has now been well measured, the agreement
between theory and experiment being approximately five percent over the measured range
(Lamoreaux, 1997). Perhaps some
variation on the Casimir cavity configuration of matter could one day be devised that will
yield  a different force on one side than on the other of some device, thus providing in
effect a ZPF-sail for propulsion through interstellar space.

\medskip\noindent
Independent of whether a propulsive force may be generated from the ZPF is the question of
whether energy can be extracted from the ZPF. This was first considered --- and found
to be a possibility in theory --- in a thought experiment published by Forward (1984).
Subsequently Cole and Puthoff (1993) analyzed the thermodynamics of zero-point energy
extraction in some detail and concluded that, in principle, no laws of thermodynamics are
violated by this. There is the possibility that nature is already tapping zero-point
energy in the form of very high energy cosmic rays and perhaps in the formation of the
sheet and void structure of clusters of galaxies (Rueda, Haisch and Cole, 1995). Another
very useful overview is the USAF Report by Forward (1996) and also the discussion of
force generation and energy extraction in Haisch and Rueda (1999).

\bigskip\centerline{\bbf THE INERTIA RESONANCE AND THE DE BROGLIE WAVELENGTH}
\medskip\noindent
Can the inertia of matter be modified?
In 1994 we, together with H. Puthoff, published an analysis using the techniques of
stochastic electrodynamics in which Newton's equation of motion, ${\bf F}=m{\bf a}$, was
{\it derived} from the electrodynamics of the ZPF (Haisch, Rueda and Puthoff [HRP], 1994).
In this view the inertia of matter is reinterpreted as an electromagnetic reaction force.
A NASA-funded research effort at
Lockheed Martin in Palo Alto and California State University in Long Beach recently led to
a new analysis that succeeded in deriving both the Newtonian equation of motion, ${\bf
F}=m{\bf a}$, and the relativistic form of the equation of motion, ${\cal F}=d{\cal
P}/d\tau$, from Maxwell's equations as applied to the ZPF (Rueda \& Haisch, 1998a; 1998b).
This extension from classical to relativistic mechanics increases confidence in the
validity of the hypothesis that the inertia of matter is indeed an effect
originating in the ZPF of the quantum vacuum. Overviews of these concepts may be found in
the previous STAIF proceedings and other conference proceedings (Haisch and Rueda, 1999;
Haisch, Rueda and Puthoff, 1998; Haisch and Rueda, 1998; additional articles are posted or
linked at http://www.jse.com/haisch/zpf.html).

\medskip\noindent
In the HRP analysis of 1994 it appeared that the crucial interaction
between the ZPF and the quarks and electrons constituting matter must be concentrated near
the Planck frequency. As discussed in the previous section (and in the appendix), the
Planck frequency is the highest possible frequency in nature and is the presumed cutoff of
the ZPF spectrum:
$\nu=(c^5/G\hbar)^{1/2}\sim 1.9 \times 10^{43}$ Hz. In contrast, the new approach involves
the assumption that the crucial interaction between the quarks and electrons constituting
matter and the ZPF takes place not at the ZPF cutoff, but at a resonance frequency. We
have now found evidence that, for the electron, this resonance must be at its
Compton frequency: $\nu=m_e c^2/h = 1.236
\times 10^{20}$ Hz. This is 23 orders of magnitude lower (hence possibly within reach of
electromagnetic technology) than the Planck frequency.

\medskip\noindent
In Rueda and Haisch (1998a, 1998b) we show that from the force associated with the
non-zero ZPF momentum flux (obtained by calculating the Poynting vector) in transit
through an accelerating object, the apparent inertial mass derives from the energy density
of the ZPF as follows:

$$m_i= {V_0 \over c^2} \int \eta(\nu) \rho_{ZP}(\nu) d\nu \ . \eqno(2) 
$$

\medskip\noindent
where $\eta(\nu)$ is a scattering parameter (see below).

\medskip\noindent
It was proposed by de Broglie that
an elementary particle is associated with a localized wave whose frequency is the Compton
frequency, yielding the Einstein-de Broglie equation:
$$h\nu_C=m_0 c^2 . \eqno(3)
$$
As summarized by Hunter (1997): ``\dots what we regard as the (inertial) mass of the
particle is, according to de Broglie's proposal, simply the vibrational energy (divided by
$c^2$) of a localized oscillating field (most likely the electromagnetic field). From this
standpoint inertial mass is not an elementary property of a particle, but rather a
property derived from the localized oscillation of the (electromagnetic) field. De Broglie
described this equivalence between mass and the energy of oscillational motion\dots as
{\it `une grande loi de la Nature'} (a great law of nature).'' The rest mass $m_0$ is
simply $m_i$ in its rest frame. What de Broglie was proposing is that the left-hand
side of eqn. (3) corresponds to physical reality; the right-hand side is in a sense
bookkeeping, defining the concept of rest mass.

\medskip\noindent
De Broglie
assumed that his
wave at the Compton frequency originates in the particle itself. An alternative
interpretation is that a particle  ``is tuned to a wave originating in the high-frequency
modes of the zero-point background field'' (de la Pe\~na and Cetto, 1996; Kracklauer,
1992). The de Broglie oscillation would thus be due to a resonant interaction with the ZPF,
presumably the same resonance that is responsible for creating inertial mass as in eq.
(2). In other words, the ZPF would be driving this $\nu_C$ oscillation.

\medskip\noindent
We therefore suggest that an elementary charge driven to oscillate at the Compton frequency
by the ZPF may be the physical basis of the $\eta(\nu)$ scattering parameter in eqn.
(2).  For the case of the electron, this would imply that $\eta(\nu)$ is a
sharply-peaked resonance at the frequency, expressed in terms of energy, $h\nu=512$ keV.
The inertial mass of the electron would physically be the reaction force due to resonance
scattering of the ZPF at that frequency.

\medskip\noindent
This leads to a surprising corollary. It can be shown (Haisch and Rueda, 1999; de la Pe\~na
and Cetto, 1996; Kracklauer, 1992) that as viewed from a laboratory frame, the standing
wave at the Compton frequency in the electron frame transforms into a traveling wave
having the de Broglie wavelength,
$\lambda_B=h/p$, for a moving electron. The wave nature of the moving electron appears to
be basically due to Doppler shifts associated with its Einstein-de Broglie resonance
frequency.

\medskip\noindent
The identification of the resonance
frequency with the Compton frequency would solve a fundamental mystery of quantum
mechanics: Why does a moving particle exhibit a de Broglie wavelength of $\lambda = h/p$?
It can be shown that if the electron acquires its mass because it is driven to oscillate at
its Compton frequency by the ZPF, then when viewed from a moving frame, a beat frequency
arises whose wavelength is precisely the de Broglie wavelength (Haisch \& Rueda, 1999; de
la Pe\~na \& Cetto, 1996). Thus within the context of the zero-point field inertia
hypothesis we can simultaneously and suggestively explain both the origin of mass and the
wave nature of matter as ZPF phenomena. Furthermore, the relative accessibility of the
Compton frequency of the electron encourages us that an experiment to demonstrate mass
modification of the electron by techniques of cavity quantum electrodynamics may soon be
feasible.

\bigskip\centerline{\bbf ACKNOWLEDGMENTS}
\medskip\noindent
We acknowledge support of NASA contract NASW-5050 for this work. 

\bigskip\centerline{\bbf REFERENCES}

{

\bigskip
\parskip=0pt plus 2pt minus 1pt\leftskip=0.25in\parindent=-.25in

Alcubierre, M.,
``The Warp Drive: Hyper-fast Travel Within General Relativity,''
{\it Class. Quantum Grav.}, {\bf 11}, L73 (1994).

Cole, D.C. and Puthoff, H.E., ``Extracting Energy and Heat from the Vacuum,'' {\it Phys.
Rev. E}, {\bf 48}, 1562 (1993).

Davisson, C.J. and Germer, L.H., ``Diffraction of Electrons by a Crystal of Nickel,'' {\it
Phys. Rev}, {\bf 30}, 705 (1927).

de la Pe\~na, L. and Cetto, A.M., {\it The Quantum Dice: An Introduction to Stochastic
Electrodynamics}, (Kluwer Acad. Publ.), (1996).

Forward, R., ``Extracting electrical Energy from the Vacuum by Cohesion of Charged
Foliated Conductors,'' {\it Phys. Rev. B}, {\bf 30}, 1700 (1984).

Forward, R.,
``Mass Modification Experiment Definition Study,''
{\it J. of Scientific Exploration}, {\bf 10}, 325 (1996).

Haisch, B. and Rueda, A., ``Reply to Michel's `Comment on Zero-Point Fluctuations and the
Cosmological Constant','' {\it Astrophys. J.}, {\bf 488}, 563 (1997).

Haisch, B., and Rueda, A.,
``The Zero-Point Field and Inertia,''
in {\it Causality and Locality in Modern Physics}, (G. Hunter, S. Jeffers and J.-P. Viger,
eds.), (Kluwer Acad. Publ.), 171, (1998).
xxx.lanl.gov/abs/gr-qc/9908057

Haisch, B., and Rueda, A.,
``Progress in Establishing a Connection Between the Electromagnetic Zero-Point Field and
Inertia.'' AIP Conference Proceedings No. 458, p. 988 (1999)
xxx.lanl.gov/abs/gr-qc/9906069

Haisch B. and Rueda, A.,
``On the Relation Between Zero-point-field-induced Inertial Mass and the Einstein-de
Broglie Formula.'' {\it Phys. Lett. A}, in press (2000)
xxx.lanl.gov/abs/gr-qc/9906084

Haisch, B., Rueda, A. and Puthoff, H.E. (HRP),  ``Inertia as a Zero-point-field Lorentz
Force,'' {\it Phys. Rev. A}, {\bf 49}, 678 (1994).

Haisch, B., Rueda, A. and Puthoff, H.E,
``Advances in the Proposed Electromagnetic Zero-point-field Theory of Inertia.''
AIAA 98-3143 (1998)
xxx.lanl.gov/abs/physics/9807023

Haroche, S. and Raimond, J.M., ``Cavity Quantum Electrodynamics,''
{\it Scientific American}, {\bf 268}, No. 4, 54 (1993)

Hunter, G.,
``Electrons and photons as soliton waves,''
in {\it The Present Status of the Quantum Theory of Light}, (S. Jeffers et al. eds.),
(Kluwer Acad. Publ.), pp. 37--44 (1997).

Kracklauer, A.F.,
``An Intuitive Paradigm for Quantum Mechanics.''
{\it Physics Essays}, {\bf 5}, 226 (1992).

Lamoreaux, S.K., ``Demonstration of the Casimir Force in the 0.6 to 6 $\mu$m Range,''
{\it Phys. Rev. Letters}, {\bf 78}, 5 (1997)

Loudon, R., {\it The Quantum Theory of Light}, chap. 1, (Oxford: Clarendon Press) (1983).

McCrea, W.,
``Time, Vacuum and Cosmos,''
{\it Q. J. Royal Astr. Soc.}, {\bf 27}, 137 (1986).

Misner, C.W., Thorne, K.S. and Wheeler, J.A., {\it Gravitation}, (Freeman and Co.), pp.
426--428 (1973).

Pfenning, M.J. and Ford, L.H.,
``The Unphysical Nature of `Warp Drive',''
{\it Class. Quant.Grav.}, {\bf 14}, 1743-1751 (1997).

Puthoff, H.E., ``Gravity as a Zero-point-fluctuation Force,'' {\it Phys. Rev. A}, {\bf
39}, 2333 (1989).

Puthoff, H.E., 
``Polarizable-Vacuum (PV) Representation of General Relativity,''
preprint, (1999). \break
xxx.lanl.gov/abs/gr-qc/9909037

Rueda, A. and Haisch, B., ``Inertia as Reaction of the Vacuum to Accelerated Motion,''
{\it Physics Lett. A}, {\bf 240}, 115 (1998a).
xxx.lanl.gov/abs/physics/9802031

Rueda, A. and Haisch, B., ``Contribution to Inertial Mass by Reaction of the Vacuum to
Accelerated Motion,'' {\it Foundations of Physics}, {\bf 28}, 1057 (1998b).
xxx.lanl.gov/abs/physics/9802030

Rueda, A., Haisch, B. and Cole, D. C., ``Vacuum Zero-Point Field Pressure Instability
in Astrophysical Plasmas and the Formation of Cosmic Voids,'' {\it Astrophys. J.}, {\bf
445}, 7 (1995).

Sakharov, A.D.,
``Vacuum Quantum Fluctuations in Curved Space and the Theory of Gravitation,''
{\it Sov. Phys.-Doklady} {\bf 12}, No. 11, 1040 (1968).

Van den Broeck, C.
``A `warp drive' with Reasonable Total Energy Requirements,''
preprint, (1999) \break
xxx.lanl.gov/abs/gr-qc/9905084 

}
\bigskip\centerline{\bbf APPENDIX: THE PLANCK FREQUENCY}
\medskip\noindent
The Planck frequency is assumed to be
the highest frequency that spacetime itself can sustain. This can be understood from
simple, semi-classical arguments by combining the constraints of
relativity with those of quantum mechanics. In a circular orbit, the acceleration is
$v^2/r$, which is obtained from a gravitational force per unit mass of
$Gm/r^2$. Letting $v \rightarrow c$, one obtains a maximum acceleration of $c^2/r=Gm/r^2$,
from which one derives the Schwarzschild radius for a given mass: $r_S=Gm/c^2$. The
Heisenberg uncertainty relation specifies that $\Delta x \Delta p \ge \hbar$, and letting
$\Delta p \rightarrow mc$ one arrives at the Compton radius: $r_C=\hbar/mc$, which
specifies the minimum quantum size for an object of mass $m$. Equating the minimum quantum
size for an object of mass $m$ with the Schwarzschild radius for that object one arrives at
a mass:
$m_P=(c\hbar / G)^{1/2}$ which is the Planck mass, i.e. $2.2 \times 10^{-5}$ g. The Compton
radius of the Planck mass is called the Planck length: $l_p=(G\hbar/c^3)^{1/2}$, i.e. $1.6
\times 10^{-33}$ cm. One can think of this as follows: Due to the uncertainty relation, a
Planck mass cannot be compressed into a volume smaller than the cube of the Planck
length. A Planck mass, $m_P$, in a Planck volume, $l_P^3$,  is the maximum density of
matter that can exist without being unstable to collapsing spacetime fluctuations:
$\rho_{P,m}=c^5/G^2\hbar$ or as an energy density,
$\rho_{P,e}=c^7/G^2\hbar$. The speed-of-light limit constrains the fastest oscillation that
spacetime can sustain to be
$\nu_P=c/l_P=(c^5/G\hbar)^{1/2}$, i.e. $1.9 \times 10^{43}$ Hz.

\bye